\documentclass[prl,twocolumn,amsmath,amssymb,superscriptaddress]{revtex4-1}
\usepackage{amsfonts}
\usepackage{amsmath}
\usepackage{amssymb}
\usepackage{graphicx}
\usepackage[T1]{fontenc}
\usepackage[utf8]{inputenc}
\usepackage{hyperref}
\usepackage[dvipsnames]{xcolor}
\usepackage{color}
\usepackage{bm}
\usepackage{comment}
\usepackage{mathtools}
\usepackage[section]{placeins}
\usepackage[capitalize]{cleveref}
\usepackage{footnote}
\usepackage{etoolbox}
\usepackage[caption=false]{subfig}

\newcommand{\q}[1]{``#1''}

\crefrangelabelformat{equation}{(#3#1#4--#5\crefstripprefix{#1}{#2}#6)}

\begin{document}
\title{Pauling entropy, metastability and equilibrium in \texorpdfstring{Dy$_2$Ti$_2$O$_7$}{Dy2Ti2O7} spin ice}

\author{S. R. Giblin} 
\affiliation{School of Physics and Astronomy, Cardiff University, CF24 3AA Cardiff, United Kingdom}
\email{giblinsr@cardiff.ac.uk}

\author{M. Twengstr{\"o}m} 
\affiliation{Department of Physics, Royal Institute of Technology, SE-106 91 Stockholm, Sweden}

\author{L. Bovo} 
\affiliation{London Centre for Nanotechnology and Department of Physics and Astronomy, University College London, 17-19 Gordon Street, London, WC1H OAH, U.K.}
\affiliation{Department of Innovation and Enterprise, University College London, 90 Tottenham Court Rd, Fitzrovia, London W1T 4TJ, UK}

\author{M. Ruminy} 
\affiliation{Laboratory for Neutron Scattering and Imaging, Paul Scherrer Institut, CH-5232 Villigen PSI, Switzerland}

\author{M. Bartkowiak} 
\affiliation{Laboratory for Neutron Scattering and Imaging, Paul Scherrer Institut, CH-5232 Villigen PSI, Switzerland}

\author{P. Manuel} 
\affiliation{ISIS Facility, Rutherford Appleton Laboratory, Chilton, Didcot, OX11 0QX, United Kingdom}

\author{J. C. Andresen}
\affiliation{Department of Physics, Ben Gurion University of the Negev, Beer Sheva 84105, Israel}

\author{D. Prabhakaran}
\affiliation{Clarendon Laboratory, Physics Department, Oxford University,Oxford, OX1~3PU, United Kingdom}

\author{G. Balakrishnan}
\affiliation{Department of Physics, University of Warwick, Coventry, CV4 7AL, United Kingdom}

\author{E. Pomjakushina} 
\affiliation{Laboratory for Scientific Developments, Paul Scherrer Institut, CH-5232 Villigen PSI, Switzerland}

\author{C. Paulsen}
\affiliation{Institut N\'{e}el, C.N.R.S - Universit\'e Joseph Fourier, BP 166, 38042 Grenoble, France}

\author{E. Lhotel}
\affiliation{Institut N\'{e}el, C.N.R.S - Universit\'e Joseph Fourier, BP 166, 38042 Grenoble, France}

\author{L. Keller} 
\affiliation{Laboratory for Neutron Scattering and Imaging, Paul Scherrer Institut, CH-5232 Villigen PSI, Switzerland}

\author{M. Frontzek} 
\affiliation{Neutron Scattering Division, Oak Ridge National Laboratory, Oak Ridge, TN, USA}

\author{S. C. Capelli} 
\affiliation{ISIS Facility, Rutherford Appleton Laboratory, Chilton, Didcot, OX11 0QX, United Kingdom}

\author{O. Zaharko} 
\affiliation{Laboratory for Neutron Scattering and Imaging, Paul Scherrer Institut, CH-5232 Villigen PSI, Switzerland}

\author{P. A. McClarty} 
\affiliation{Max Planck Institute for the Physics of Complex Systems, N{\"o}thnitzer Str. 38, 01187 Dresden, Germany}

\author{S. T. Bramwell}
\affiliation{London Centre for Nanotechnology and Department of Physics and Astronomy, University College London, 17-19 Gordon Street, London, WC1H OAH, U.K.}

\author{P. Henelius} 
\affiliation{Department of Physics, Royal Institute of Technology, SE-106 91 Stockholm, Sweden}

\author{T. Fennell} 
\affiliation{Laboratory for Neutron Scattering and Imaging, Paul Scherrer Institut, CH-5232 Villigen PSI, Switzerland}
\email{tom.fennell@psi.ch}

\begin{abstract}  

Determining the fate of the Pauling entropy in the classical spin ice material Dy$_2$Ti$_2$O$_7$ with respect to the third law of thermodynamics has become an important test case for understanding the existence and stability of ice-rule states in general. The standard model of spin ice {--} the dipolar spin ice model {--} predicts an ordering transition at $T\approx 0.15$ K, but recent experiments by Pomaranski \emph{et al.} suggest an entropy recovery over long time scales at temperatures as high as $0.5$ K, much too high to be compatible with theory. Using neutron scattering and specific heat measurements at low temperatures and with long time scales ($0.35$ K$/10^6$ s and $0.5$ K$/10^5$ s respectively) on several isotopically enriched samples we find no evidence of a reduction of ice-rule correlations or spin entropy. High-resolution simulations of the neutron structure factor show that the spin correlations remain well described by the dipolar spin ice model at all temperatures. Further, by careful consideration of hyperfine contributions, we conclude that the original entropy measurements of Ramirez \emph{et al.} are, after all, essentially correct: the short-time relaxation method used in that study gives a reasonably accurate estimate of the equilibrium spin ice entropy due to a cancellation of contributions. 
\end{abstract}

\maketitle

The properties of ice-rule states, such as those of water ice~\cite{pauling35,giauque36} and spin ice~\cite{harris97,bramwell98,ramirez99}, provide a strong contrast with the conventional paradigm of condensed matter. Instead of broken symmetry, entropy that vanishes in accord with the third law, exponentially decaying correlations and wavelike excitations, one finds Coulomb phase correlations~\cite{fennell09}, finite entropy~\cite{pauling35,ramirez99},  and point-like fractional excitations (monopoles)~\cite{ryzhkin05,castelnovo08}.  The mapping between the hydrogen bonding network and spin configurations~\cite{tajima82,bramwell98}, and the resultant identical residual (Pauling) entropy~\cite{ramirez99} are cornerstones of spin ice physics, posing fundamental questions such as how a realistic Hamiltonian can lead to practical evasion of the third law and whether the entropic state is metastable?  Because the low-temparature dynamics of spin-ice depends on a vanishing number of thermally excited monopoles, relaxation becomes very slow at low temperatures~\cite{giauque36,snyder04}, and sensitivity to sample variations is enhanced~\cite{revell12,sala14}; both effects may mask the true equilibrium state. While the third law ground state of water ice can be accessed by doping that increases dynamics~\cite{tajima82}, the fate of the residual entropy in the spin ice Dy$_2$Ti$_2$O$_7$ ~\cite{ramirez99} is not known.  Because of these experimental challenges, the problem of third law ordering in ice-type systems may best be addressed by a careful collaboration of experiment and theory, designed to accurately model the system and to extrapolate properties beyond the experimental range.

\begin{figure*}
    \centering{\resizebox{1\hsize}{!}{\includegraphics{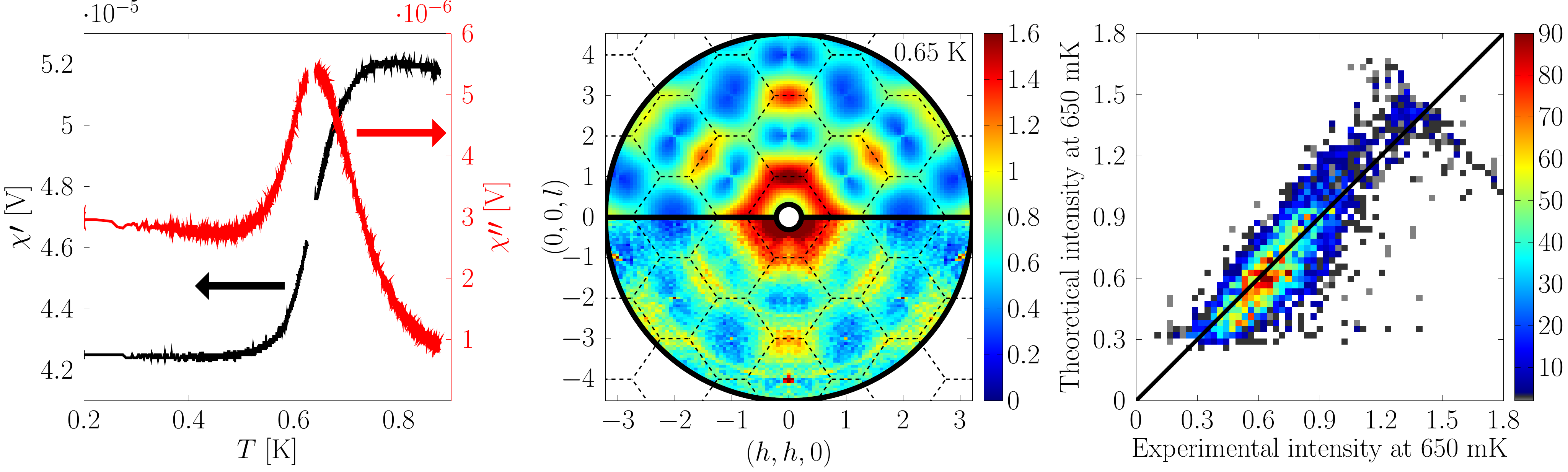}}}
    
    \vspace{-8.5mm}
    \hspace{-165mm}
    \subfloat[\label{susc_cal}]{}
    
    \vspace{-8.5mm}
    \hspace{-50mm}
    \subfloat[\label{NS_PSI}]{}
    
    \vspace{-8.5mm}
    \hspace{70mm}
    \subfloat[\label{int_map}]{}
    
    \caption{Measurement of spin correlations with accurately equilibrated spin temperature of $0.65$ K. (a) Example in-situ susceptibility measurements (here susceptibility $\chi$ is a linear function of measured voltage). (b) The measured (lower hemisphere) and simulated (upper hemisphere) neutron structure factors at $0.65$ K are in close agreement. (c) Comparison of measured and calculated intensities (the color scale indicates the number binned points, with 11518 in total).} 
    
\end{figure*}

The spin ice state of Dy$_2$Ti$_2$O$_7$~\cite{harris97,bramwell98,ramirez99} is a consequence of frustration arising from the competition between the Ising-like crystal field anisotropy~\cite{bertin12,ruminy16}, exchange, and dipolar interactions~\cite{denhertog00,melko01}.  These ingredients can be described by a simple classical spin Hamiltonian {--} the dipolar spin ice model (DSM)~\cite{denhertog00,melko01}, which has been refined for many years~\cite{ruff05,yavo08,henelius16}.  Such theoretical investigations suggest that the dipolar spin ice states have a small band width that  eventually leads to an ordering transition and the recovery of the residual entropy~\cite{denhertog00,melko01,ruff05,yavo08}.  However, several specific heat studies have found no indication of ordering or reduction in the residual entropy, so the Hamiltonian parameters are required to suppress any such process below $T\sim 0.3$ K.  In fact, the population of monopole excitations becomes very small below the so-called freezing temperature, $T_{\rm f}\approx0.65$ K in Dy$_2$Ti$_2$O$_7$, where experimental timescales diverge exponentially.  Recently, Pomaranski~\emph{et al.}~\cite{pomaranski13} combined very long equilibration times ($\approx 10^5$ s) with very accurate temperature measurements to show that the specific heat apparently increases below $0.5$ K. This experimental {\it tour de force} caused considerable excitement: is the residual entropy recovered at a much higher temperature than predicted, indicating an insufficiency in the DSM that might even allow an alternative, non-classical ground state~\cite{mcclarty14, henelius16}?  

Since performing well equilibrated measurements much below $T_{\rm f}$ is challenging, a well parametrized Hamiltonian model is important to allow for predictions of the low-temparature properties. Diffuse neutron scattering data, a measure of the spin-spin correlation function, is well suited to directly test the Hamiltonian. However, the lowest temperature data used previously were obtained at 0.3 K with relatively short equilibration times ($\sim10^3$ s)~\cite{fennell04}, a procedure that may be reasonably questioned~\cite{henelius16} in light of the subsequently discovered long equilibration times. In this letter, we describe neutron scattering measurements in the static approximation, designed to measure the spin and lattice temperatures of the sample, in-situ. We verify the spin ice Hamiltonian with well-controlled equilibration at $T=0.65$ K, and, by monitoring the spin system in-situ for $0.35<T<0.4$ K over a period $>2\cdot 10^6$ s we demonstrate that there is no evidence of any change in correlations or emergence of diffraction peaks in this temperature and time window. Specific heat measurements on various isotopically enriched Dy$_2$Ti$_2$O$_7$ samples show that the system comes to thermodynamic equilibrium in this range, and we find no evidence for a recovery of the Pauling entropy at 0.5 K.

The main sample studied by neutron scattering was a $1.4$ g $^{162}$Dy$_2$Ti$_2$O$_7$ cuboid originally described in Ref. [\onlinecite{fennell04}].  For this work the sample was reannealed in oxygen, and the SXD (ISIS, UK)~\cite{sxd} and TRiCS (PSI, Switzerland)~\cite{trics} diffractometers were used to confirm that no structural diffuse scattering indicative of oxygen defect clusters~\cite{sala14} was present~\cite{supp}.  Magnetic diffuse neutron scattering experiments were performed on the WISH (ISIS)~\cite{wish} and DMC (PSI) ~\cite{dmc} diffractometers.  At DMC Helmholtz coils reduced stray fields below $0.1$ $\mu$T. WISH has a stray field of~$100$ $\mu$T.  The crystal was clamped in a copper goniometer with continuous thermal path to the mixing chamber (MC) of the dilution fridge.  The clamp extends along $3/4$ of the length of the sample, with a protruding few mm surrounded by a small a.c. susceptometer that was thermalised by a de-oxygenated copper braid to the mixing-chamber end of the goniometer. The coil set was surrounded by neutron absorbing cadmium. RuO$_2$ thermometers were attached to the goniometer, close to the sample and the MC (12 cm apart). Due to the well defined frequency dependence of the spin relaxation in $^{162}$Dy$_2$Ti$_2$O$_7$~\cite{snyder04,matsuhira11,yara12} the a.c. susceptometer can be used to measure the effective spin temperature, simultaneously with measurements of the lattice temperature by the sample thermometer.  Fig.~\ref{susc_cal} shows a representative relative susceptibility measurement, taken at 0.5 Hz, which shows a peak in the imaginary part centered around 0.64 K.  From calibration measurements it can be estimated that the coil set induces a small (0.02 K) heating effect as the protruding part of the sample is not well thermally coupled, but that the lattice and spin temperature are well coupled.  

\begin{figure*}
    \centering{\resizebox{1\hsize}{!}{\includegraphics{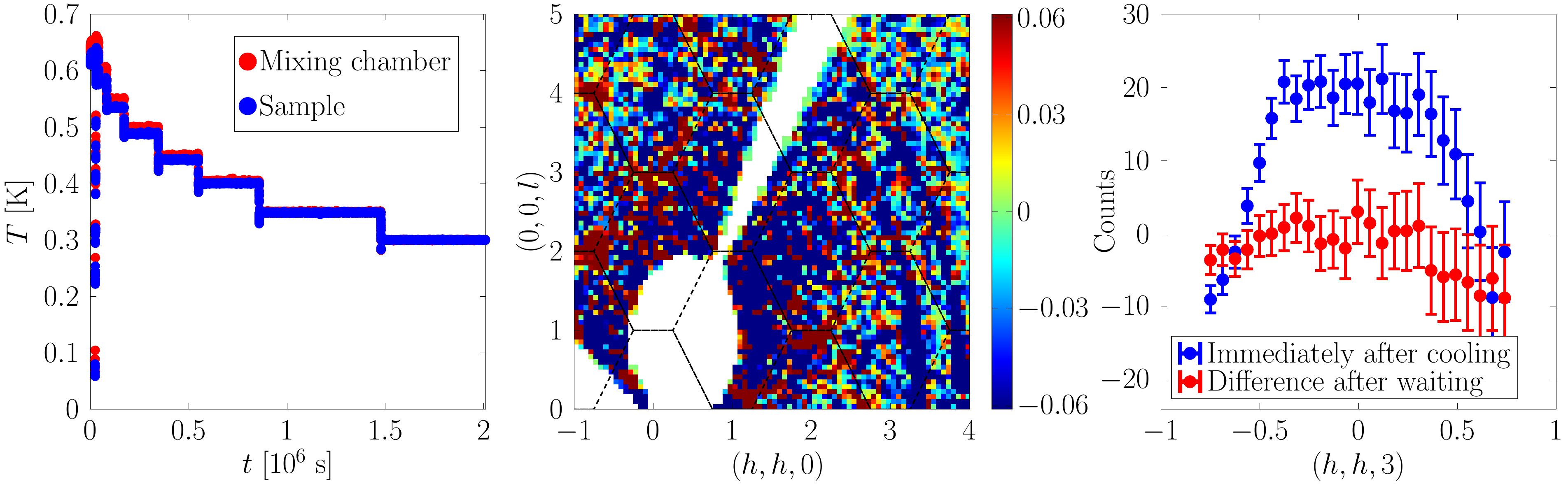}}}
    
    \vspace{-9.5mm}
    \hspace{-165mm}
    \subfloat[\label{temp_wait}]{}
    
    \vspace{-8.5mm}
    \hspace{-55mm}
    \subfloat[\label{diff_map}]{}
    
    \vspace{-8.5mm}
    \hspace{80mm}
    \subfloat[\label{line_cut}]{}
    
    \caption{Slow cooling and equilibration at 0.35 K.  The sample was cooled by a protocol similar to Ref.~\cite{pomaranski13} (a).  Comparing a rapidly cooled measurement at 0.35 K with the slow cooled measurement by taking the difference of the structure factors shows no difference between the two (b).  This is highlighted by comparing cuts through the quenched and slow cooled data and difference map (c).} 
    
\end{figure*}

To model the neutron structure factor we use the dipolar spin ice Hamiltonian
\begin{align}\label{dip_ham}
	\mathcal{H} =& \sum_{i>j}J_{ij}\bm{S}_i\cdot\bm{S}_j\nonumber\\
	&+Da^3\sum_{i>j}\left[\frac{\bm{S}_i\cdot\bm{S}_j}{r_{ij}^3}-3\frac{\left(\bm{S}_i\cdot\bm{r}_{ij}\right)\left(\bm{S}_j\cdot\bm{r}_{ij}\right)}{r_{ij}^5}\right],
\end{align}
where $\bm{S}_i$ are the spin vectors, $a$ is the nearest neighbour distance, $r_{ij}$ the distance separating particle $i$ and $j$, $D$ is the dipolar constant~\cite{yavo08} and $J_{ij}$ is a matrix describing the coupling strength between particle $i$ and $j$. In this study we used the $g^+${--} dipolar spin ice model ($g^+${--}DSM) exchange parameters ($J_1=3.41$ K, $J_2=-0.14$ K, $J_{3a}=-0.030$ K , $J_{3b}=-0.031$ K)~\cite{bovo18,supp}. Using a parallel Monte Carlo code that exploits the symmetry of the dipolar interactions~\cite{ewald21}, we could reach system sizes of $L=16$ ($65536$ spins), an order of magnitude larger and with improved resolution for $S\left(\mathbf{Q}\right)$ than the majority of studies~\cite{henelius16}.  We used periodic Ewald boundary conditions~\cite{ewald21} and  a loop algorithm~\cite{melko01} to speed up
equilibration. Using parallel tempering we found the ordering temperature~\cite{melko01} of the $g^+${--}DSM to be $T_{\rm c}=0.15 $ K.

\begin{figure*}
    \centering{\resizebox{1\hsize}{!}{\includegraphics{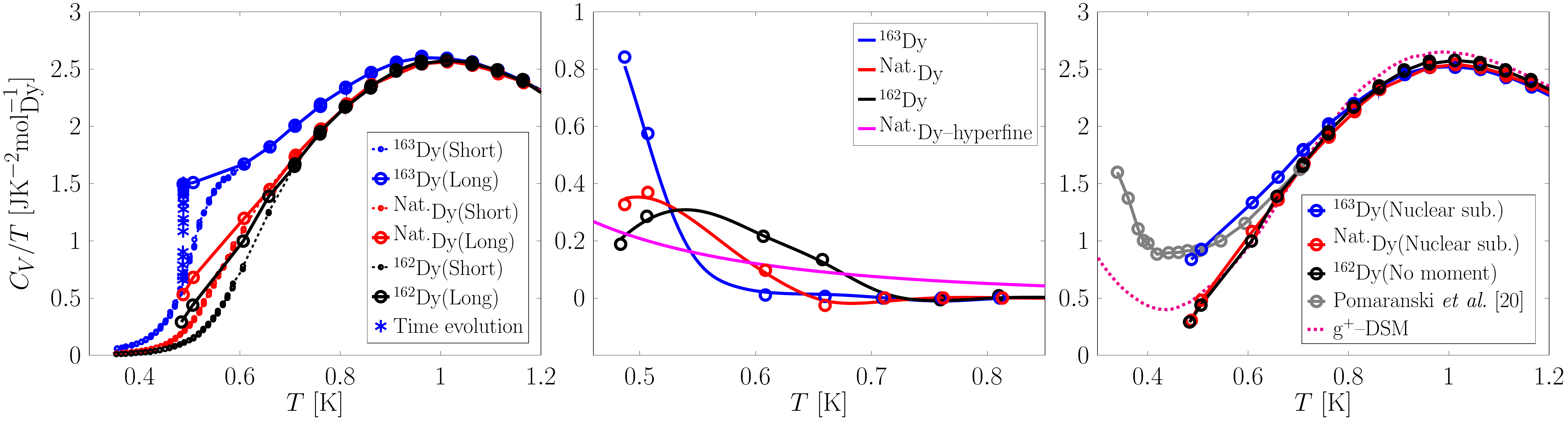}}}
    
    \vspace{-9.5mm}
    \hspace{-165mm}
    \subfloat[\label{cv_long_short}]{}
    
    \vspace{-8.5mm}
    \hspace{-50mm}
    \subfloat[\label{cv_diff}]{}
    
    \vspace{-8.5mm}
    \hspace{65mm}
    \subfloat[\label{cv_pom_nuk}]{}
    
    \caption{Specific heat of $^{162}$Dy, $^{\textup{Nat.}}$Dy and $^{163}$Dy samples of Dy$_2$Ti$_2$O$_7$. (a) Long and short-time measurements. (b) Difference between long and short-time measurements along with the calculated nuclear specific heat for the $^{\textup{Nat.}}$Dy sample. (c) Equilibrium electronic specific heat for all samples, as well as model calculation and previous results from Pomaranski~\emph{et al.}~\cite{pomaranski13} (adjusted for the correct hyperfine contribution~\cite{henelius16}).}
    
\end{figure*}

We first verify that the $g^+${--}DSM parameters describe the diffuse neutron scattering data well at $T=0.65$ K, the lowest temperature where we expect no equilibration issues (well below existing data at 1.3 K~\cite{henelius16}). At this temperature the frequency dependence of the susceptibility allows us to verify the spin temperature of the sample, and simultaneously measure the diffuse neutron scattering (DMC).   Fig.~\ref{NS_PSI} shows the experimental and simulated neutron structure factors for $T=0.65$ K. The best fit of the theory to experiment was made  by scaling the experimental data using a linear point-wise function and then performing an RMS-minimisation in a region including the $(0,0,3)$ and $(3/2,3/2,3/2)$ peaks.  As can be seen from the color map, and the comparison of experimental and calculated intensities (Fig.~\ref{int_map}), the experimental and theoretical data match well at this temperature. The calculation captures the essence of the experimental data: the intensity and relative weight at $\mathbf{Q}=(0,0,3)$ and $\mathbf{Q}=(3/2,3/2,3/2)$ and the zone-boundary scattering, with no  unexpected intensity at any wave vector.  Similar agreement is obtained for $T=1.3$ K (see Ref.~\cite{supp}).  We therefore expect that Dy$_2$Ti$_2$O$_7$ is well described by the g$^+$-DSM at lower temperature, with an ordering transition around $T_{\rm c}=0.15 $ K.

To examine the lower temperature equilibration, and seek any behavior beyond the $g^+-$DSM, we needed to cool below $T_{\rm f}$ ensuring that  we understood the thermal state of our sample during the measurements. Using a frequency of $0.5$ Hz, we could monitor the cooling of the sample for $0.5$ K $<T<0.9$ K.  The lattice temperature and heat flows in/out of the sample can be monitored with the sample thermometer, and we saw that the spin temperature and lattice temperature were in equilibrium with each other, and with the MC. There were identifiable sources of experimental heating below 0.5 K - either by operating the susceptometer below 0.5 K, or below 0.3 K from the neutron beam.  The neutron beam heating effect depends on the neutron flux and thermal coupling, so is experiment specific. Measurements using the thermometer on the MC alone would not observe this heating, demonstrating the importance of in-situ monitoring for poor thermal conductors. 

We performed two experiments with different cooling protocols, the results of which are identical.  We discuss the second, whose cooling protocol was longer than that of Pomaranski~\emph{et al.}~\cite{pomaranski13,kycia_cool}.  Fig.~\ref{temp_wait} shows the cooling of the thermometers attached to the goniometer and MC.  These experiments were performed at WISH, cooling during an ISIS accelerator shutdown.  To avoid self-heating effects, we made this measurement with the susceptometer switched off, having established that significant heat flows in/out of the sample can be recorded by the sample thermometer. The estimated equilibration time and beam heating constraints meant that our target temperature was 0.35 K.  To look for changes in the correlations with long equilibration times, we performed a difference measurement at 0.35 K. After initial standard cooling of the sample from 0.65 K to 0.35 K (in some minutes), a map was recorded. We then warmed the sample back to 0.65 K and followed the cooling protocol shown in Fig.~\ref{temp_wait}, and recorded a second map at 0.35 K~\cite{supp}.  As can be seen in the difference map, Fig.~\ref{diff_map}, no statistically significant difference can be detected, indicating that the previous low-temparature data, and analyses based on it~\cite{fennell04,yavo08,henelius16}, are trustworthy. A cut across $\mathbf{Q}=(0,0,3)$ for the difference of the two measurements is compared to similar cuts for the unsubtracted data sets in Fig.~\ref{line_cut}. As can be expected, within the statistical quality of our data, there is no peak in the difference and the two measurements are identical.  The $g^+-$DSM parameters also reproduce this diffuse scattering well (see Ref.~\cite{supp}).

To further check for consistency we performed specific heat measurements by the relaxation technique, paying particular attention to the long relaxation time~\cite{pomaranski13} and the nuclear hyperfine interaction~\cite{henelius16}. We used a Quantum Design Physical Properties Measurement System with $^3$He insert, modified to allow relaxation times from $10^1$ s and $10^5$ s, largely covering the timescales probed by Pomaranski~\emph{et al.}~\cite{pomaranski13}. We used three isotopes: an off-cut  of the $^{162}$Dy$_2$Ti$_2$O$_7$ neutron sample, with no nuclear spin ($I=0$); an enriched $^{163}$Dy$_2$Ti$_2$O$_7$ sample with nuclear spin $I=5/2$; and  a natural abundance of isotopes sample ($^{\textup{Nat.}}$Dy), containing $^{161}$Dy and $^{163}$Dy, both with nuclear spin $I = 5/2$, fractions 0.19, 0.249 respectively, and the remainder with $I=0$. The hyperfine energy is sufficiently small that the hyperfine specific heat $C_{\rm H}$ may be taken as equal to its high temperature value $C_{\rm H} = a/T^2$ where $a = 0$, $0.026$, $0.076$ JK$^{-1}$mol$^{-1}_{\textup{Dy}}$ for $^{162}$Dy,  $^{\textup{Nat.}}$Dy and $^{163}$Dy respectively (here $a = R (1/3)I(I+1)\epsilon^2$, $\epsilon$ is the effective energy separation of hyperfine levels and $R$ is the gas constant). Adjusting for the incorrect nuclear spin (hyperfine) specific heat in Ref.~\cite{pomaranski13} ($ = 0.0052/T^2$) does not account for all the upturn in the specific heat~\cite{henelius16} (see also Fig.~\ref{cv_pom_nuk}).

In Fig.~\ref{cv_long_short} we show the specific heat for the three samples measured as a function of temperature for two extreme relaxation times ($12$ s and $89\times 10^4$ s). A comparison of time-dependent measurements at our base temperature (not shown) suggests that all samples reach complete thermodynamic equilibrium at $t \sim 10^5$ s, as found in Ref. \cite{pomaranski13}. We clearly see that increasing the timescale leads to an increase in the specific heat, as noticed by Pomaranski \emph{et al.}~\cite{pomaranski13}. Furthermore the nuclear contribution is clearly visible at low temperature, with $^{162}$Dy ($I=0$) having the lowest specific heat, and  $^{163}$Dy ($I=5/2$) the highest.

The difference between the long- and short-time measurements for each sample is shown in Fig.~\ref{cv_diff}, along with the hyperfine contribution for $^{\textup{Nat.}}$Dy. The short-time measurements remain in equilibrium to a lower temperature when there are more nuclear spins present. This indicates that the hyperfine energy levels provide additional relaxation paths for the electronic spins, and also shows that the nuclear relaxation rate is on the same order as the electronic relaxation rate for $T\geq 0.55$ K. 

Finally, we show the equilibrium electronic specific heat for the three samples, after the nuclear contributions has been subtracted in Fig.~\ref{cv_pom_nuk}. Our main result is the comparison of the $^{\textup{Nat.}}$Dy and the $^{162}$Dy sample. In this case, subtraction of the hyperfine contribution projects the two curves very accurately onto each other at all temperatures revealing the equilibrium electronic specific heat (plus a negligible phonon contribution). We have high confidence in this result: the $^{162}$Dy sample was removed from the larger sample that has been extensively characterized by neutron scattering, while the standard methods used to prepare the $^{\textup{Nat.}}$Dy should render it relatively free of defects and impurities, as our results imply. In addition, the specific heat of all three individual crystals, from two different crystal growers, collapse above 0.8 K, which is not the case for many previous measurements~\cite{pomaranski13}.  This experimental equilibrium curve is well, but not perfectly, matched by the g$^+${--}DSM.  These slight discrepancies are due to the fact that the original g{--}DSM parameters~\cite{yavo08} were adjusted to short-time specific heat data that was slightly higher~\cite{higa03}. Further fine-tuning of the parameters is possible, but not the aim of this study.  Our experiments establish that there is no upturn in the equilibrated electronic specific heat above $T=0.5$ K, a conclusion fully consistent with theory and our neutron measurements. 

Our results suggest that the most likely cause for the upturn observed by Pomaranski \emph{et al.}~\cite{pomaranski13} is random disorder, in agreement with Ref.~\cite{henelius16}. Magnetic defects introduce localized energy levels, which will increase the specific heat per magnetic ion, rather than diminish it, so spin ice samples with the largest equilibrium (or long time) specific heat tend to be the most defective, as our study implies. 

Reconsidering now the short-time measurements in Fig.~\ref{cv_long_short}, we note that the short-time curve for the $^{\textup{Nat.}}$Dy sample is not so far away from the final equilibrium curve, the long-time curve for the $^{162}$Dy sample. Hence the uncorrected  short-time results of Ramirez \emph{et al.}~\cite{ramirez99} and other authors are reasonable estimates of the equilibrium electronic specific heat. The reason for this is a cancellation of terms: adding the long-time contribution and subtracting the nuclear part leads to only a small net change in the specific heat, since these terms are roughly equal, see Fig.~\ref{cv_diff}. Also notable is that the short-time measurements of all three samples converge below 0.4 K. This suggests that the nuclear relaxation time becomes much greater than the electronic one at low temperature, and that the hyperfine specific heat is not visible in a short-time measurement for $T\leq0.45$ K~\cite{bertin02}.

In conclusion, this letter addresses a specific case of the more general question: how can we know the third law ground state of ice-type systems, whose dynamics depend on a vanishing number of pointlike excitations (monopoles)? When the monopole density becomes very small, extrinsic defects and disorder become important: in the case of water ice they are thought to provide sufficient dynamics to locate the ground state~\cite{tajima82}  but there seems to be no comparable mechanism available in spin ice.   In $^{162}$Dy$_2$Ti$_2$O$_7$ we have carefully equilibrated the sample at 0.65 K and 0.35 K, demonstrating the value of measuring the temperature of both the spin and lattice baths when characterizing such systems using neutron scattering.  By confirming the accuracy of the dipolar spin ice model in this range, we support the “monopole fluid” picture of spin ice~\cite{castelnovo08} and the interesting theories and experiments that arise from it~\cite{revell12,paulsen16,kaiser15,kaiser18}.  We predict the recovery of the Pauling entropy at lower temperatures, and our work highlights the experimental temperature and time windows that would have to be accessed to detect effects beyond the standard model of spin ice.

\begin{acknowledgments}
We thank J. Kycia and D. Pomaranski for sharing their data of the cooling protocol in Ref.~\cite{pomaranski13}, M. J. P. Gingras for useful discussions and suggestions, the late S.N. Fisher for a selection of calibrated thermometers and the ISIS sample environment team. The simulations were performed on resources provided by the Swedish National Infrastructure for Computing (SNIC) at the Center for High Performance Computing (PDC) at the Royal Institute of Technology (KTH). M.T. and P.H. are supported by the Swedish Research Council (2013-03968), M.T. is grateful for funding from Stiftelsen Olle Engkvist Byggmästare (2014/807), M. R. was supported by the SNSF (Schweizerischer Nationalfonds zur F\:oorderung der Wissenschaftlichen Forschung) (Grant No. 200021\_140862), and L.B. was supported by The Leverhulme Trust through the Early Career Fellowship program (ECF2014-284). G.B. thanks EPSRC, UK for funding through Grant EP/M028771/1

S.R.G./M.R./M.B./P.M./T.F./M.T./P.A.M. performed neutron scattering experiments (with L.K./M.F. at PSI).
S.R.G./M.B./C.P./E.L. designed, constructed and calibrated the susceptometer and thermometry.
M.T./S.R.G./M.R./P.H. analysed neutron scattering data and compared to simulations.
M.R./E.P./T.F./S.R.G./P.M. reannealed and characterized the $^{162}$Dy$_2$Ti$_2$O$_7$ crystal with O.Z./S.C.C..
L.B./S.T.B. developed/carried out the specific heat measurements and their analysis.
M.T./J.C.A./P.H. developed/carried out the Monte Carlo simulations.
D.P. and G.B. grew the crystals ($^{163}$Dy and $^{\textup{Nat.}}$Dy, $^{162}$Dy respectively).
S.R.G./M.T./S.T.B./P.H./T.F. wrote the paper with input from all authors.

\end{acknowledgments}

\newpage

\renewcommand{\thetable}{S\arabic{table}}
\renewcommand{\thefigure}{S\arabic{figure}}
\renewcommand{\bibnumfmt}[1]{[S#1]}
\renewcommand{\citenumfont}[1]{S#1}
\setcounter{figure}{0}

\onecolumngrid

\subsection*{\fontsize{12}{15}\selectfont Supplementary material for \q{Pauling entropy, metastability and equilibrium in Dy$_2$Ti$_2$O$_7$ spin ice}}

\subsection{Model parameters}
\begin{figure}[htb!]
    \centering{
    \resizebox{0.5\hsize}{!}{\includegraphics{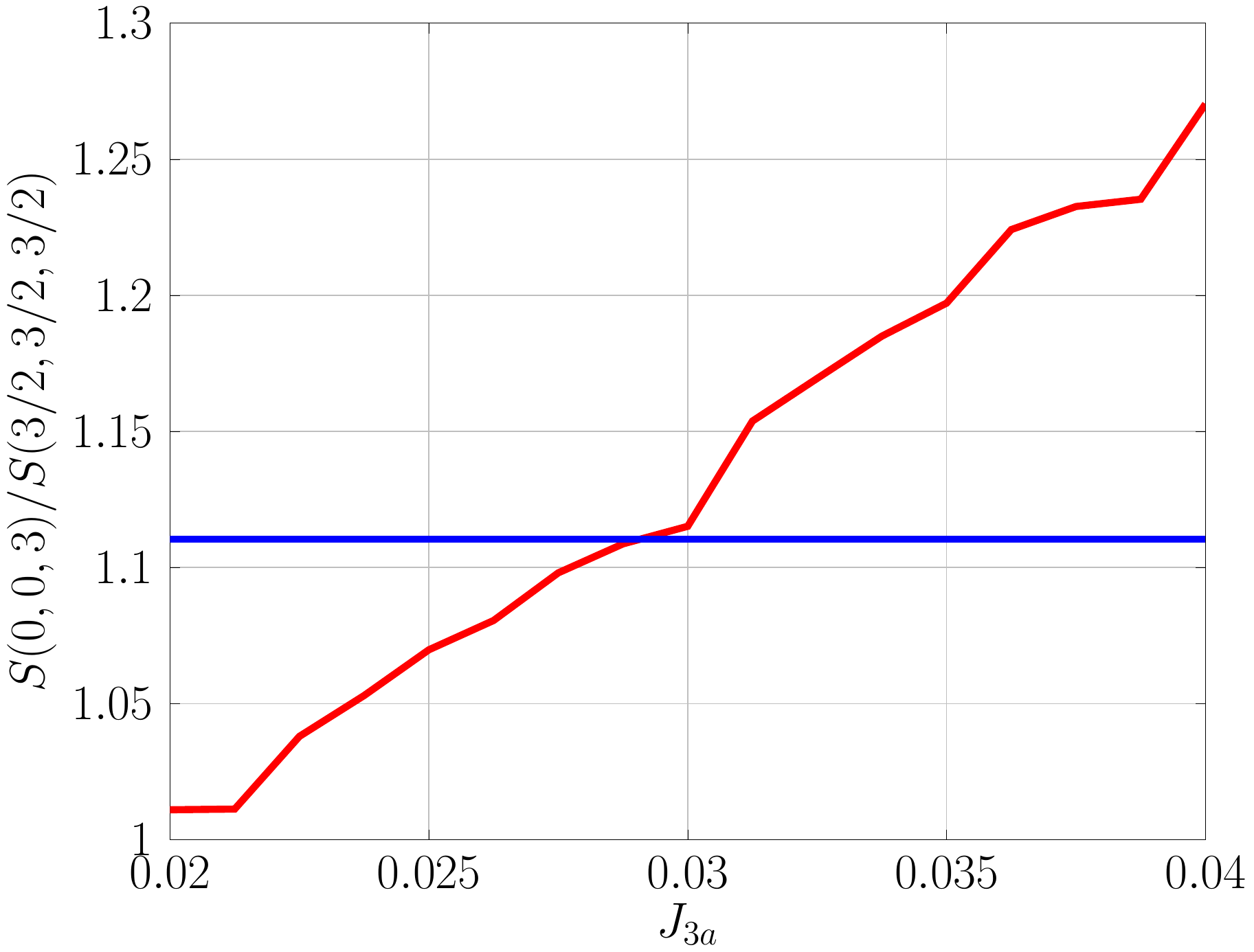}}}
    \caption{The ratio of the intensities at $\mathbf{Q}=(0,0,3)$ and $\mathbf{Q}=(3/2,3/2,3/2)$ vs $J_{3a}$ values along the minimum RMS valley at $T=0.65$ K. The blue line corresponds to the experimental value of the ratio at $T=0.65$ K.}
    \label{rel_int}
\end{figure}

While it may be challenging to determine a complete set of exchange parameters for a material based solely on measurements of the neutron structure factor, it can be very useful in conjunction with other measurements. A parameter set which describes a number of Dy$_2$Ti$_2$O$_7$ measurements well is the general dipolar spin ice model (g-DSM), with dipolar constant $D=1.322$ K, and exchange constants $J_1=3.41$ K, $J_2=-0.14$ K, $J_{3a}=J_{3b}=0.025$ K~\cite{Syavo08}. However, a recent study~\cite{Sbovo18} finds that these parameter values do not describe the peak in the intrinsic magnetic susceptibility $\chi T/C$ accurately. Retaining $J_1=3.41$ K, $J_2=-0.14$ K, a quantitative  description of $\chi T/C$ leads instead to the relation $J_{3b}=-0.80J_{3a}+0.056$ K. From this relation the point closest to the g{--}DSM model was chosen, to define the g$^+${--}DSM, ($J_1=3.41$ K, $J_2=-0.14$ K, $J_{3a}=-0.030$ K, $J_{3b}=-0.031$ K). In order to investigate this choice of third nearest neighbors we calculated the ratio $S(0,0,-3)/S(3/2, 3/2, 3/2)$ while varying $J_{3a}$ according to the relation $J_{3b}=-0.80J_{3a}+0.056$ K and compared these ratios to our experimental value at $T=650$ mK. As shown in Fig.~\ref{rel_int} the experimental and calculated ratios match when  $J_{3a}\approx-0.030$ K, which was the value chosen in Ref.~\cite{Sbovo18}. Our results provides further experimental justification for this parameter choice and underscores the usefulness of using neutron scattering data to determine weak further neighbour exchange parameters. 

\subsection{Numerical equilibration}

The resolution in reciprocal space is inversely proportional to the number of particles used in the simulation. In this study, we reach system sizes of $65536$ spins, which is $8$ times larger than in commonly reported simulations. Large system sizes pose an intricate problem due to the difficulty of reaching equilibrium in a theoretical sense. Equilibration of $S\left(\mathbf{Q}\right)$ was established by comparing the difference of maps by successively increasing the number of steps according to the following procedure: calculate the average neutron map, along with the standard deviation, for  a large set of independent maps using $2^{N}$ MC steps. Next, double the simulation length and again calculate the average map, along with the standard deviation, for the same number of  independent maps, using $2^{N+1}$ MC steps. Finally, calculate the standard deviation of the two average maps generated using $2^{N}$  and $2^{N+1}$ MC steps. Repeat this process by successively increasing the value of $N$ until the standard deviation calculated from the average maps, generated using different numbers of MC steps, is smaller than the standard deviation within the maps generated in the longer simulation. The production chart is then produced by using $2^{N+1}$ equilibration steps and $2^{N+2}$ measuring steps. We find that $N=15$ is sufficient for reaching equilibrium.

This method can be viewed as a version of logarithmic binning. The main reason for choosing this scheme instead of logarithmic binning is that we find $S\left(\mathbf{Q}\right)$ to equilibrate unevenly fast for different $\mathbf{Q}$ values.

\subsection{Neutron scattering maps}

\begin{figure*}[htb!]
    \centering{\resizebox{0.9\hsize}{!}{\includegraphics{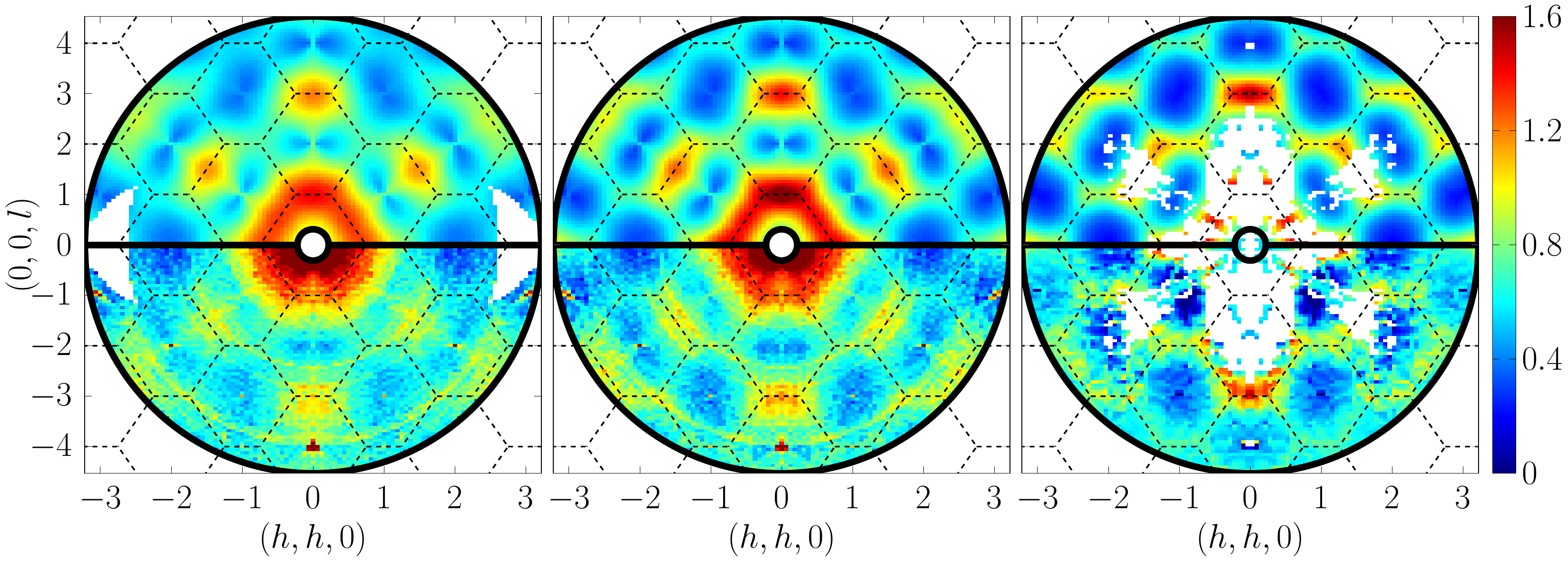}}}
    
    \vspace{-9.5mm}
    \hspace{-150mm}
    \subfloat[\label{1300mK_map}]{}
    
    \vspace{-8.5mm}
    \hspace{-50mm}
    \subfloat[\label{650mK_map}]{}
    
    \vspace{-8.5mm}
    \hspace{45mm}
    \subfloat[\label{350mK_map}]{}
    
    \caption{Neutron scattering maps at $1.3$ K, $0.65$ K and $0.35$ K from left to right. The lower semicircles contains experimental data, while the upper semicircles contain theoretically calculated maps. While we believe that the entire maps at 650 and 1300 mK, taken at PSI, are accurate, the data at 350 mK is  noisy in certain regions due to the flux profile of the time of flight diffractometer used at ISIS. The area around $\mathbf{Q}=(0,0,-3)$ features low noise, but the intensity around $\mathbf{Q}=(3/2,3/2,-1/2)$ is dominated by counting noise.}
    
\end{figure*}

The experimental and theoretical scattering maps for 1.3, 0.65 and 0.35 K are shown in~\cref{1300mK_map,650mK_map,350mK_map}. The higher temperature experimental maps were taken on DMC (PSI)~\cite{Sdmc} and the 0.35 K data were taken on WISH (ISIS)~\cite{Swish}. The experimental data taken by the two instruments was normalized by scaling the 0.65 K data taken on WISH to the 0.65 K data from DMC. The resulting scale parameter was used to scale the intensity of the 0.35 K WISH data, clearly displaying the expected increase of intensity as the temperature is lowered. Due to the geometry of the WISH experiment, data was collected for a single orientation optimized for scattering around the (0,0,3) region of intensity. As a consequence, statistically reliable data is limited to a restricted part of the reciprocal space, and in~\cref{350mK_map} we have removed the regions where errors due to Poisson statistics dominate.

\subsection{Neutron scattering at long equilibration time}

\begin{figure*}[htb!]
    \centering{\resizebox{0.6\hsize}{!}{\includegraphics{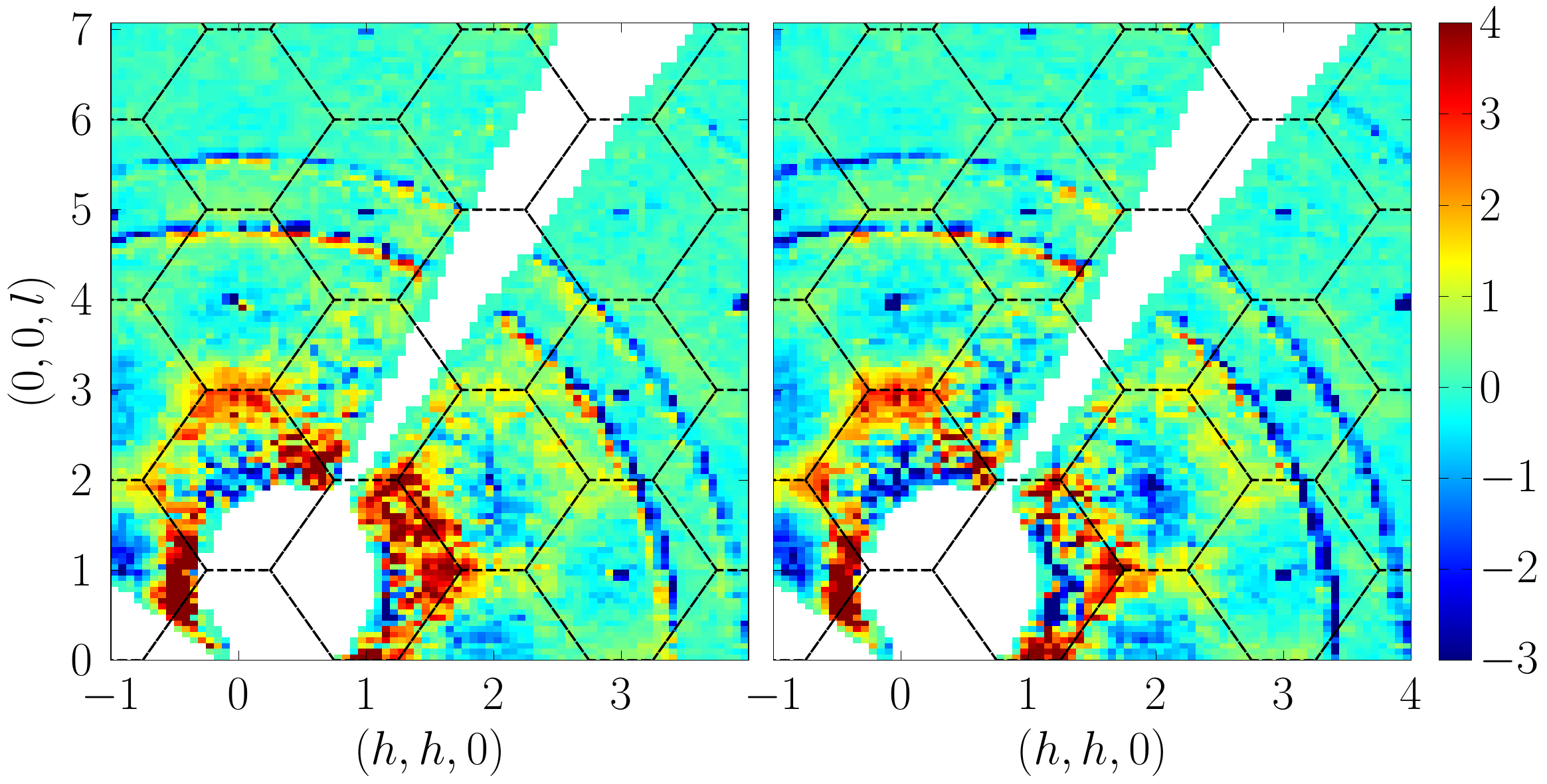}}}
    
    \vspace{-9.5mm}
    \hspace{-100mm}
    \subfloat[\label{fast_350mK_map}]{}
    
    \vspace{-8.5mm}
    \hspace{-5mm}
    \subfloat[\label{slow_350mK_map}]{}
    
    \caption{Neutron scattering maps at $0.35$ K taken immediately after cooldown (a) and after waiting (b).}
    
\end{figure*}

In~\cref{fast_350mK_map,slow_350mK_map} we show the experimental map taken immediately after cooling and waiting respectively, on the WISH diffractometer (ISIS). In both cases a high temperature ($12$ K) background map, measured during the experiment, was subtracted from the data to clearly show the structure associated with spin ice. Clearly visible in both are powder diffraction rings which originate from the copper clamp holding the sample in place.  A normalization factor of $1.02$ between the data recorded before and after the long equilibration is required to make the intensity of these features equal.  This is  because of the slight change in response of the neutron monitor over the timescale of the experiment.  In Fig.~2b the copper rings have been removed by the subtraction.

\subsection{Oxygen annealing}
    In Fig. S4, we show the diffuse scattering at large wave vectors of our $^{162}$Dy$_2$Ti$_2$O$_7$ crystal after oxygen annealing.
    Significant diffuse scattering in this plane was shown to be a signature of oxygen defects in titanate pyrochlores in Ref.~\cite{Ssala14}. After annealing, no diffuse scattering of the type described in Ref.~\cite{Ssala14} can be detected, suggesting that our sample is relatively defect free. The data were measured on the SXD instrument at ISIS~\cite{Ssxd}.

\begin{figure}[htb!]
    \centering{\resizebox{0.4\hsize}{!}{\includegraphics{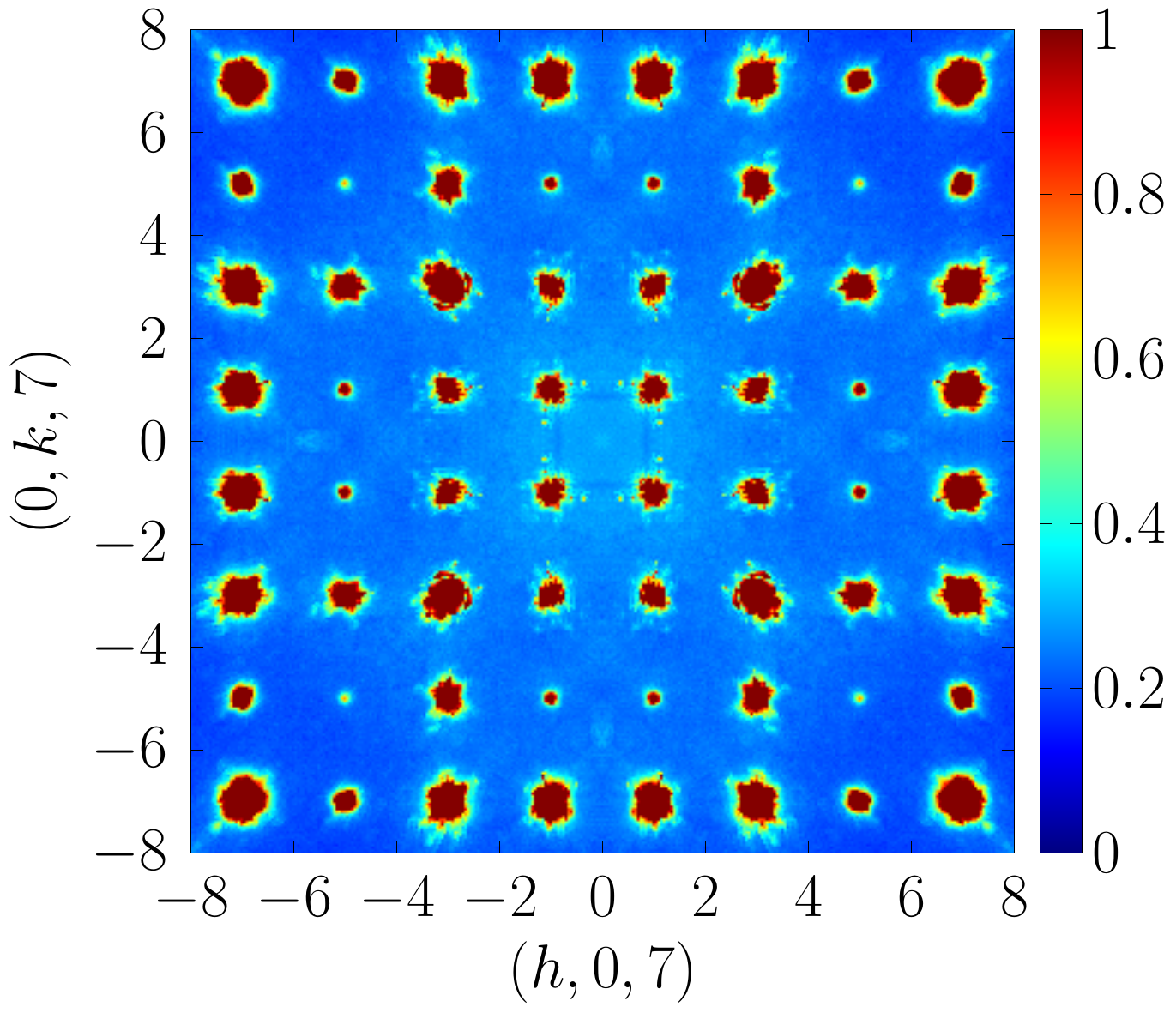}}}
    \label{SXDa}
    \caption{The diffuse neutron scattering in the $\left(h,k,7\right)$ plane after the Dy$_2$Ti$_2$O$_7$ crystal was annealed in oxygen. The scattering shows almost no contamination from defects.}
\end{figure}

\end{document}